\documentclass[manuscript,sigconf,nonacm,screen]{acmart}

\AtBeginDocument{%
  \providecommand\BibTeX{{%
    \normalfont B\kern-0.5em{\scshape i\kern-0.25em b}\kern-0.8em\TeX}}}

\usepackage{multirow}
\usepackage{makecell}

\newcommand\colH[1]{\multicolumn{1}{c}{\textbf{#1}}}

\newcommand{\shortquote}[1]{``\emph{#1}''}

\newcommand{\emj}[1]{{\color{red} Eunice: #1 }}
\newcommand{\cyrus}[1]{{\color{green} Cyrus: #1 }}

\newcommand{\markupRed}[1]{%
 {#1}%
}

\begin{document}



\title{Dreamcrafter: Immersive Editing of 3D Radiance Fields Through Flexible, Generative Inputs and Outputs}

\author{Cyrus Vachha}
\affiliation{%
    \institution{UC Berkeley}
    \city{Berkeley}
    \state{CA}
    \country{USA}
}
\email{cvachha@berkeley.edu}

\author{Yixiao Kang}
\affiliation{%
    \institution{UC Berkeley}
    \city{Berkeley}
    \state{CA}
    \country{USA}
}
\email{yixiao\_kang@berkeley.edu}

\author{Zach Dive}
\affiliation{%
    \institution{UC Berkeley}
    \city{Berkeley}
    \state{CA}
    \country{USA}
}
\email{zach\_dive@berkeley.edu}

\author{Ashwat Chidambaram}
\affiliation{%
    \institution{UC Berkeley}
    \city{Berkeley}
    \state{CA}
    \country{USA}
}
\email{ashwatc@berkeley.edu}

\author{Anik Gupta}
\affiliation{%
    \institution{UC Berkeley}
    \city{Berkeley}
    \state{CA}
    \country{USA}
}
\email{anik.gupta@berkeley.edu}

\author{Eunice Jun}
\affiliation{%
    \institution{UCLA}
    \city{Los Angeles}
    \state{CA}
    \country{USA}
}
\email{emjun@cs.ucla.edu}

\author{Björn Hartmann}
\affiliation{%
    \institution{UC Berkeley}
    \city{Berkeley}
    \state{CA}
    \country{USA}
}
\email{bjoern@eecs.berkeley.edu}

\renewcommand{\shortauthors}{Vachha et al.}



\begin{abstract}
Authoring 3D scenes is a central task for spatial computing applications. Competing visions for lowering existing barriers are (1) focus on immersive, direct manipulation of 3D content or (2) leverage AI techniques that capture real scenes (3D Radiance Fields such as, NeRFs, 3D Gaussian Splatting) and modify them at a higher level of abstraction, at the cost of high latency. We unify the complementary strengths of these approaches and investigate how to integrate generative AI advances into real-time, immersive 3D Radiance Field editing.
We introduce Dreamcrafter, a VR-based 3D scene editing system that: (1) provides a modular architecture to integrate generative AI algorithms; (2) combines different levels of control for creating objects, including natural language and direct manipulation; and (3) introduces proxy representations that support interaction during high-latency operations. 
We contribute empirical findings on control preferences and discuss how generative AI interfaces beyond text input enhance creativity in scene editing and world building. For videos and additional materials visit our project page: \url{https://dream-crafter.github.io/}

\end{abstract}








\begin{teaserfigure}
  \includegraphics[width=\textwidth]{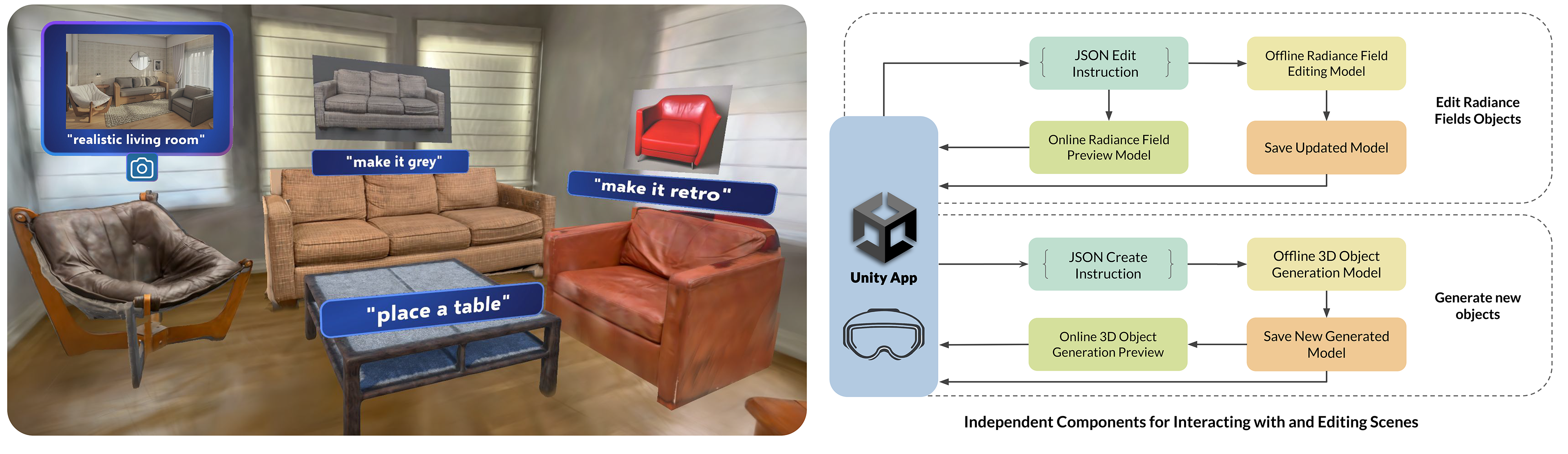}
  \caption{System Overview: (Left) View of edits, proxy generations, and spatial annotations applied while scene editing. (Right) Pipeline overview of system including a Unity app interfacing with online and offline modules to edit scenes for editing radiance fields, creating new objects, and creating 2D stylized renders.}
  \Description{(Left) - Visual 3D VR scene depicting the living room of a house. There are various objects such as a couch, chair, table, etc. spread out across the room, each accompanied by a spatial annotation representing the edit information inputted by the user. The couch has an annotation reading “make it grey”, followed by a generated 2D grey version of the couch displayed above the 3D asset. The chair has the edit instruction “make it retro” along with a retro-styled 2D image above. The table has the instruction “place a table”, and depicts a rendered 3D table on the floor. (Right) - Architecture diagram titled Independent Components for Interacting with and Editing Scenes, containing three components: Unity App, Edit Radiance Fields Objects, and Generate new objects in the scene. Edit Radiance Fields Objects contains the following workflow: Unity app feeds into JSON Edit Instruction, which feeds into Offline Radiance Field Editing Model, feeding into Save Updated Model, then Online Radiance Field Preview, and finally back to Unity App. JSON Edit Instruction also directly feeds into Online Radiance Field Preview. Generate new objects in the scene contains the following workflow: Unity app feeds into JSON Create Instruction, which feeds into Offline 3D Object Generation Model, feeding into Save New Generated Model, feeding into Online 3D Object Generation Preview, and finally back to Unity App. JSON Create Instruction also feeds into Online 3D Object Generation Preview.}
  \label{fig:teaser}
\end{teaserfigure}

\maketitle
\section{Introduction}
Spatial computing applications such as Augmented and Virtual Reality rely on 3D content and scenes. Thus, creating appropriate tools for authoring and editing 3D content has been a long-standing key challenge for HCI researchers. Traditionally, mesh-and-texture-based approaches have been used to author 3D content. Various research efforts to introduce better editing techniques notwithstanding (e.g.,~\cite{igarashi-teddy,bae-sketch}), the expertise hurdle to create and modify 3D content in this way has been high, generally leaving such authoring to a small number of expert users. 

One avenue to lower the authoring barrier has been to embrace authoring in VR (e.g. Google Tiltbrush~\cite{tiltbrush}), where direct 3D input is possible through VR controllers (or gestures)  in an immersive environment. This approach decreases the gulf of execution~\cite{hutchins1985direct} inherent in prior approaches to modeling 3D content using 2D input devices.

More recently, two additional developments hold the promise of reducing authoring burdens. First, novel approaches for representing 3D scenes based on radiance fields (e.g., NeRFs~\cite{mildenhall2020nerf} and 3D Gaussian Splatting~\cite{kerbl3Dgaussians}) allow for straightforward capture of photorealistic environments from real scenes using common cameras, instead of having to model objects from scratch. Second, generative AI developments have introduced novel ways of editing 3D scenes like radiance fields at higher levels of abstraction, e.g. through text instructions (as in Instruct-NeRF2NeRF~\cite{instructnerf2023}). While offering the ability to edit at a semantic level rather than a lower geometry level, such techniques also tend to be compute-intensive and not yet amenable to run in realtime. 

The different approaches---rapid direct manipulation on the one hand and high-level instruction-based editing on the other hand---recall long-standing arguments in the HCI community on the benefits of direct control vs. delegation~\cite{shneiderman1997direct}.
In this paper, we investigate if it is possible to unify the complementary strengths of real-time, immersive editing on the one hand, and generative AI-based approaches to high-level scene editing (with high latency) on the other hand under a common interaction framework. 
\markupRed{We believe that generative AI can enable new forms of interactions through natural language and could lower the barrier to entry for users with limited 3D modeling experience and could make it easier for users to express creative ideas, while making it faster or easier to prototype for experienced users.}

We introduce Dreamcrafter, a Virtual Reality 3D content generation and editing system assisted by generative AI.\footnote{dream-crafter.github.io} The core idea behind Dreamcrafter is to use direct manipulation for spatial positioning and layout; and leverage generative AI for editing style and appearance of objects. Because generative AI edits are unlikely to run in real-time, Dreamcrafter introduces rapid proxy representations, e.g. using a 2D diffusion model to create a stand-in image for a longer-running 3D generative task.  Dreamcrafter enables both 2D (image) and 3D output. 


Dreamcrafter makes three technical contributions: (1) Scene editing system with photo-realistic scene representations. We use radiance fields (3D gaussian splatting) instead of traditional mesh-based representations. (2) Modular architecture. This enables the system to continuously integrate state of art generative AI models and leverage both 2D and 3D proxy representations. (3) Flexibility in scene editing. A combination of voice prompts with natural language instructions and sculpting using primitives gives both general and advanced users extensive flexibility.

We chose to make Dreamcrafter a VR immersive editor since it affords interactions such as voice input and direct manipulation which are more natural in VR and allow for spatial 3D input.
Radiance fields are intrinsically realistic and immersive, and are well suited to be viewed in VR, making their creation within VR representative of a user’s experience. \markupRed{Radiance fields are being applied to a variety of applications such as in geospatial systems~\cite{cesium}, gaming~\cite{lumalabs}, social experiences~\cite{metahorizon}, e-commerce~\cite{amazon}, and education~\cite{niedermayr2024novel}.}  
There is rising industry interest in systems with immersive radiance fields such as Meta Hyperscape~\cite{metahyperscape}, Varjo Teleport~\cite{varjoteleport}, Niantic Scaniverse ~\cite{scaniverse} and Gracia AI~\cite{graciavr} which envision using radiance fields (specifically gaussian splats) for social, entertainment, and industrial applications. However, this rise in immersive radiance field usage reveals a gap in studying HCI systems and interactions tailored for editing and creating them.
Recent interfaces for generative media, such as Luma Dream Machine~\cite{lumalabs} and Runway ML~\cite{runwayml}, enable users to articulate creative visions through 2D representations like images and video, often using natural language instructions for refinement and conceptualization.
With Dreamcrafter, we aim to extend this paradigm into 3D by imagining an analogous spatial editor enabling users to engage in immersive world-building, translating similar interactions such as using natural language instructions while facilitating exploration of generative outputs akin to current systems. \markupRed{Traditional workflows for 3D modeling are time consuming and expensive to create large detailed scenes. With Dreamcrafter, users with minimal 3D modeling experience can create and edit 3D scenes for general applications which could be applied to game development, interior design, or virtual production for filmmaking.}

We investigate how users decide between different levels of control over a scene and how they use proxy representations through a first-use study with seven participants. Using Dreamcrafter, participants could either (i) generate entire objects using AI or (ii) first construct 3D objects using pre-defined shapes (i.e., spheres, cubes, etc.) and then stylize the constructions using generative AI. While participants created more objects using the former interaction, they felt more in control with the latter interaction. Regardless of generation approach, participants found the proxy previews useful for scene composition. 


\section{Background}
We give a brief overview of Radiance Fields (NeRFs and Gaussian Splatting) and generative image models. 
\paragraph{Radiance Fields.} Recent years have seen a move from traditional 3D graphics using meshes and geometries to more photorealistic rendering techniques, such as Neural Radiance Fields (NeRFs) \cite{mildenhall2020nerf} and Gaussian Splatting \cite{kerbl3Dgaussians}. Radiance fields are 3D representations of scenes or objects, as a function of radiance given position and view direction, that can exhibit photorealistic view dependent effects. NeRFs are 3D representations that optimize a volumetric 3D scene as a radiance field using a neural network trained on a set of images. 3D Gaussian Splatting is akin to NeRFs. The main difference is that Gaussian Splatting uses 3D Gaussians to support faster training and rendering via differentiable rasterization for high-quality real-time visualizations. These techniques have been shown to be highly effective at modeling details with realistic lighting, shadowing, and surfaces for real-world captures. And, with the increase in applications requiring 3D content, these models can be effectively used to quickly capture and create assets.

\paragraph{Generative Image Models.} 
Stable Diffusion \cite{rombach2022highresolution} is a deep learning model for synthesizing, or generating, images from text inputs using a diffusion model. ControlNet \cite{zhang2023adding} is a network architecture enhancement to text-to-image models to condition the model on an input image, generating stylized outputs given a text prompt. 



\section{Related Work}
The most related prior work falls into three areas: 1) novel 3D scene representations and tools for using them 2) generative scene building systems and 3) creation systems in VR. We review each area in turn.

\subsection{Generating and editing novel 3D representations using Radiance Fields}
Several recent rendering techniques build upon NeRFs \cite{mildenhall2020nerf} and 3D Gaussian Splatting (3DGS) \cite{kerbl3Dgaussians}. For example, LERFs \cite{lerf2023}, or Language Embedded Radiance Fields use CLIP embeddings \cite{radford2021learning} to allow users to query a NeRF using natural language to determine regions of interest. ConceptGraphs \cite{conceptgraphs} uses a similar technique with CLIP embeddings but processes a more traditional 3D representation of point clouds rather than NeRFs. These developments have indicated the importance of object-centric labeling and editing in the systems that are built and have guided our design of editing components of 3D scenes in VR. By focusing on radiance fields, a 3D representation that doesn't enable very low level manipulations like meshes, we hope to lower the barriers to a wide range of 3D representations (such as video), tasks, and applications in the future. 

While effective, these 3D representations are difficult for end users to create, manipulate, and use. \markupRed{Recent work in HCI literature such as SharedNeRF \cite{10.1145/3613904.3642945} demonstrates the enhanced benefits of using photorealistic 3D representations for showcasing objects that would be harder to reconstruct with traditional methods, but its system is unable to directly edit or manipulate the NeRF.} Additional efforts to make NeRFs more approachable have included consumer facing systems such as Luma AI \cite{lumalabs}, and research friendly APIs such as Nerfstudio \cite{Tancik_2023} and Instant-NGP \cite{mueller2022instant}. Instruct-NeRF2NeRF~\cite{instructnerf2023} allow users to provide as input a text prompt and an existing NeRF and output a new NeRF stylized according to the text prompt, relying on 2D text and image conditioned diffusion models such as InstructPix2Pix \cite{brooks2022instructpix2pix}. We interface with a Instruct-NeRF2NeRF like method to allow users to edit their 3D scenes and objects. DreamFusion \cite{poole2022dreamfusion} also allows users to build a NeRF or mesh from a text prompt. 

Approaches to interactive editing of radiance fields are emerging. NeRFShop \cite{NeRFshop23} allows selecting, transforming, or warping NeRF objects in a single scene with real-time feedback; however it lacks generation capabilities. GaussianEditor \cite{chen2023gaussianeditorswiftcontrollable3d} presents a web based interface for 3DGS including object generation, but crucially does not offer real-time proxies, making edits visible in 10-15 minutes. Neither offer immersive interfaces.

\subsection{Generative Scene Building Systems and Interfaces}
Previous work has explored how to build editing interfaces for generative AI models for 2D images or 3D meshes. WorldSmith \cite{10.1145/3586183.3606772} generates 2D scenes composed of multiple text prompts and blends across the generated images tiles. Similar to Dreamcrafter, it allows users different abstractions of editing such as to generate images via prompting or sketching but offers neither immersive interactions nor 3D results. Text2Room\cite{hoellein2023text2room} generates a fixed 3D room given a single text prompt. Dreamcrafter allows editing of existing 3D scenes and generation of individual objects. "What's the Game, then?"~\cite{10.1145/3654777.3676358} demonstrates capabilities of LLMs in creating scene functionality at runtime. Recent startups such as World Labs \cite{worldlabs} demonstrate text and image to 3D scene systems using generative image and video models and are currently exploring interfaces for editing and creation tools. \markupRed{Recent commercial systems such as Dream Machine~\cite{lumalabs}, Midjourney~\cite{midjourney}, and Adobe Firefly~\cite{adobefirefly} combine multi-modal image and text inputs for generative image and video generation with an iterative design process, similar to Dreamcrafter, but those systems are non-immersive, constrained to the 2D generation space and lack spatially persistent scene creation tools.}

\subsection{Creation Systems in VR}

There is a long history of creativity tools and developing systems in XR.   
3DM \cite{Butterworth19923DMAT} laid the groundwork by presenting a 3D modeling system operated via a 6-DoF mouse, offering a novel way to interact with digital objects in three-dimensional space. Building on this, ISAAC \cite{isaac} introduced scene editing within Virtual Environments, allowing for a more intuitive and immersive design process. Coninx et al. investigated hybrid 2D and 3D editing~\cite{coninx1997hybrid}. CaveCAD\cite{cavecad} demonstrates an intuitive immersive 3D modeling system. SculptUp\cite{sculptup}, a system for freeform virtual sculpting of organic shapes, enables artists and designers to conceptualize and iterate on their creations in an intuitive manner that closely mirrors the physical sculpting process. 
Furthermore, Google's TiltBrush \cite{tiltbrush} allows creators to paint with virtual light and textures, extending the canvas beyond the limits of traditional media. Similarly, VR games like Dreams \cite{dreams}, Figmin XR \cite{figmin}, and Horizon Worlds \cite{metahorizon} have provided valuable insights into user interaction models, offering a glimpse into how VR can facilitate complex design tasks while maintaining user-friendly interfaces. 
\markupRed{Immersive prototyping tools in various applications have also been explored. In a review of the landscape of XR tools\cite{10.1145/3558192}, Nebeling emphasizes the importance of context and story in XR design as well as the need for tools that complement existing workflows and support rapid prototyping. Henrikson et al. \cite{10.1145/2984511.2984539} explore how XR tools can be applied to the filmmaking process through prototyping with digital storyboards. Pronto \cite{10.1145/3313831.3376160} demonstrates an XR system for rapid prototyping AR experiences with proxy drawings.} Han et al.~\cite{Han2023-uf} demonstrate the next steps in HCI design and interaction with virtual environments by increasing accuracy and range of physical gesture recognition, an approach that lends itself to more natural and user-friendly interaction with the surrounding virtual environment. 
Several projects explore immersive scene editing for related domains. Flowmatic~\cite{10.1145/3379337.3415824} explores arranging interactive elements. 360proto~\cite{10.1145/3290605.3300826} enables prototyping VR and AR interfaces through paper mockups. 360proto can help visually arrange scenes through layering but has limited editing capabilities. Neither focuses on interactions for generative AI.
There is recent work in exploring how generative AI can be used for VR scene creation. Manesh et al. ~\cite{10.1145/3643834.3661547} discuss how people form natural language instructions when creating scenes in VR. VRCopilot~\cite{10.1145/3654777.3676451} introduces an immersive scene creator that uses generative AI for designing room layouts. These systems explore the use of LLMs in assisting scene creation (positioning and layouts), but Dreamcrafter leverages generative AI for designing objects in scenes with support for non-traditional 3D representations like gaussian splats. Davis et al. explore methods for collaborative AI systems to navigate design solutions for generating 3D objects from GANs in VR ~\cite{10.1145/3450741.3465260}. In comparison, Dreamcrafter provides a full 3D editor incorporating additional generative image models for staging scenes instead of primarily exploring generations of objects.

More recently, researchers have begun to explore the incorporation of generative AI in virtual environments. For example, the Large Language Model for Mixed Reality (LLMR) framework \cite{de2023llmr} leverages Large Language Models (LLMs) and the Unity game engine for real-time creation and modification of interactive Mixed Reality experiences, showcasing the potential of LLMs to facilitate intuitive and iterative design in mixed reality applications. 
Style2Fab\cite{faruqi2023style2fab} also demonstrates the ability of generative models in personalized 3D model generation. The Dynamics-Aware Interactive Gaussian Splatting System \cite{jiang2024vr} also enables the creation of animated and interactive experiences within virtual reality settings.



Our system leverages generative AI and natural language to assist in 3D scene editing in virtual environments, but prior and concurrent works don't aim to create creativity tools leveraging radiance fields.
\section{Design Goals}
Based on our review of related work, we identified a gap in research on interaction techniques and systems for working with emerging radiance field technologies and generative AI. Current generative AI interfaces are predominantly focused on image or video outputs and rely heavily on text-based prompts, with some recent advancements exploring multimodal, chat-based interactions. However, these systems remain limited in their ability to support the creation and manipulation of 3D content. With Dreamcrafter, our goal is to design a 3D editor capable of not only supporting radiance fields but also accommodating future implicit 3D representations, such as undiscovered formats for 3D video or world models ~\cite{NEURIPS2018_2de5d166}. This approach addresses the need for tools that can evolve alongside advancements in generative AI and 3D representations, enabling users to interact with and edit complex 3D environments more effectively.

We aimed to develop tools for a design space that supports interactions with high levels of abstraction, generalizable across various 3D representations.
Our design goals were conceived by analyzing recent advancements in computer vision and graphics research, where we identified limitations in new 3D representations (of NeRFs and gaussian splats) and generative AI models for images/3D and worked to design interaction techniques that could seamlessly integrate these representations and interactions into a 3D editor framework.
We also examined the affordances of generative models, such as natural language instructions and diffusion-based 2D and 3D generative models, to inform our design process. Drawing inspiration from existing 2D generative AI interfaces like those used for image or video generation, we focused on interactions such as natural language prompts and multimodal (image + text) inputs that allow users to refine their outputs. Based on this, our approach bridges the gap between emerging generative AI capabilities and intuitive, flexible interaction techniques for future implicit or explicit 3D representations. \markupRed{Making Dreamcrafter an immersive VR editor allows us to integrate more natural interactions like speech and helps the user design the environment while experiencing it.} 

Therefore, we formulated the following design goals:
\begin{itemize}
    \item \textbf{Focus on creating and editing radiance field objects in VR.} We want to support users in populating 3D scenes with radiance field objects. This may involve updating objects already in the scene or creating completely new objects. Users may want to generate new objects based on very specific prompts which may not exist in a catalog or want to create objects with more controllability in a specific style to match other objects in the scene. \markupRed{Generating objects is faster than searching through a library and is ideal for adding scene-specific objects not already available especially when building stylized scenes.}
    \item \textbf{Enable both direct manipulation and instruction-based editing.} Users may prefer different levels of control for various scene editing tasks. For example, users may want to directly manipulate objects for detailed edits while preferring natural language instructions for larger scale edits. Users should have access to both.
    
    \item \textbf{Offer modular architecture to allow integration of future generative AI advances.} An important aim of Dreamcrafter is to provide users with state-of-the-art 3D object editing and generative models, so a modular framework is necessary. In the fast-paced field of computer vision, SOTA models are introduced rapidly, and to address this, we designed our system to easily integrate new methods and support a range of interactions, including generative models that translate low-fidelity inputs to high-fidelity outputs such as through image re-stylization.

    \item \textbf{Preserve real-time interaction regardless of the latency of editing operations.} For real-time scene editing, users should not be hindered  by the system's latency. In the event that a process cannot be performed online, users should have access to previews of the edits they have made to the VR environment. This is motivated by the limitations of existing computer vision methods for generation and editing, which are often computationally intensive and time-consuming
\end{itemize}



  \begin{figure}[h]
    \centering
    \resizebox{0.9\linewidth}{!}{%
      \includegraphics{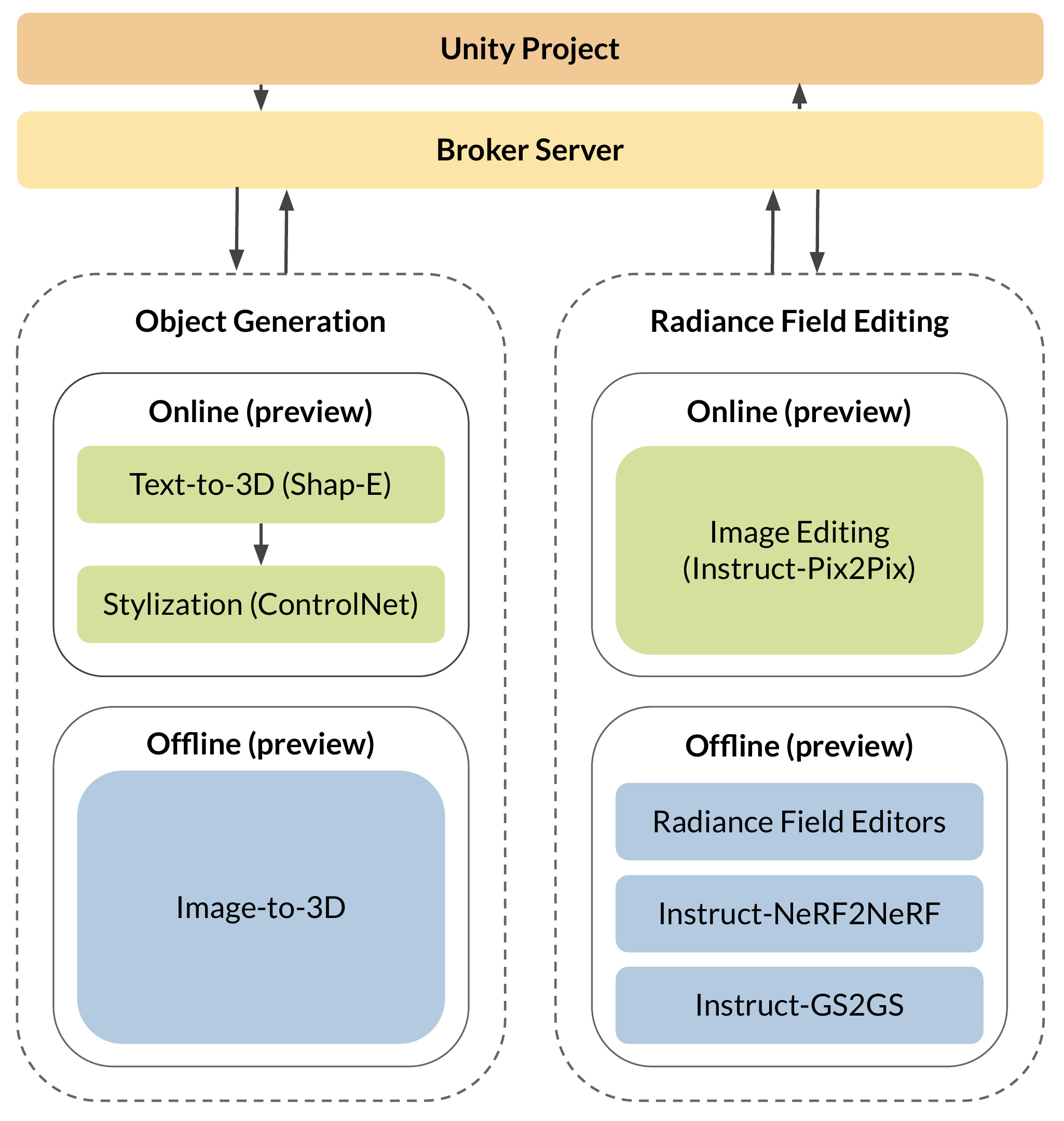}
    }
    \label{fig:sysOverview}
    \caption{\textbf{Dreamcrafter system overview.} Modules processing pipeline: The Unity project sends API calls to the broker server to run instructions from specific generation modules and their outputs get sent back to the Unity project. Online modules are run for previewing generations, and offline modules are run after editing is complete.}
    \Description{Four components: Unity Project, Broker Server, Object Generation, Radiance Field Editing. Unity Project serves as the backbone, feeding into and out of the Broker Server. Broker Server independently communicates with Object Generation and Radiance Field Editing independently. Object Generation contains: Online (Preview) component connecting Text-to-3D (Shap-E) to Stylization (ControlNet), and Offline component containing Image to 3D Model. Radiance Field Editing contains: Online (Preview) component containing Image Editing (Instruct-Pix2Pix), and Offline component combining Radiance Field Editors, Instruct-NeRF2NeRF, and Instruct-GS2GS.}
  \end{figure}

\section{System Design and Implementation}
Dreamcrafter provides an interface to edit and generate radiance field objects using generative AI-enabled tools. Dreamcrafter supports different levels of user control and gives real-time proxy representations to preview time-consuming edits and introduces new workflows leveraging image diffusion models (i.e., Stable Diffusion). Users can select fixed regions in space or existing objects in the scene to apply spatial annotations. Existing or pre-captured radiance field objects can be added to the scene via an object menu. Generations and edits can be re-done or deleted. Each type of edit and module is designed in the framework to be interchangeable and modular allowing new types of interactions to be added in the future, or replace existing ones. Spatial annotations are added to objects or spaces that are assigned edits with corresponding proxy representations based on edit instructions. \autoref{fig:teaser} shows spatial annotations applied in a scene.

\subsection{Key interactions}
Dreamcrafter supports four interactions for moving, editing, and generating new radiance field objects. 


\begin{figure}[h]
\centering
\includegraphics[width=\linewidth]{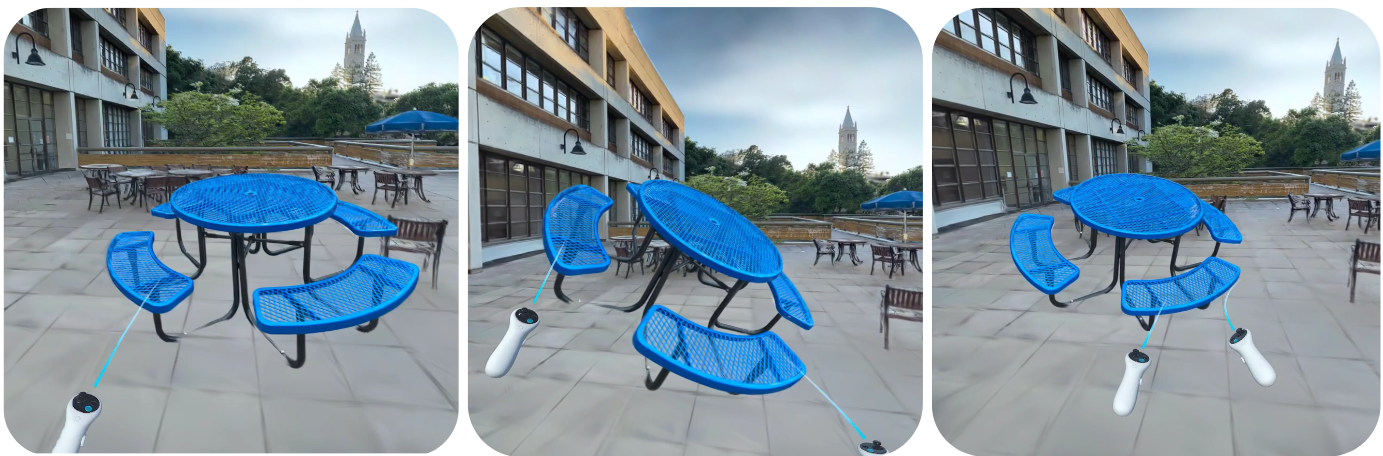}
\caption{\textbf{Object transformations and direct manipulations:} (Left) Positioning object in the scene (Center) Rotating object. (Right) Scaling object}
\label{fig:moveObjs}
\Description{User is holding VR controllers to manipulate the scene, in a simulated outdoor environment. They depict rotations, translations, and scaling of a 3D table bench asset in space.}
\end{figure}

\subsubsection{Move objects}
Users can move objects (generated or radiance field based) with spatial manipulations with hand movements and VR controls. Objects can be positioned, rotated, or scaled within the scene. Physics can be applied to help align the objects or stack generated objects. \autoref{fig:moveObjs} illustrates this interaction.


\begin{figure}[h]
\centering
\includegraphics[width=\linewidth]{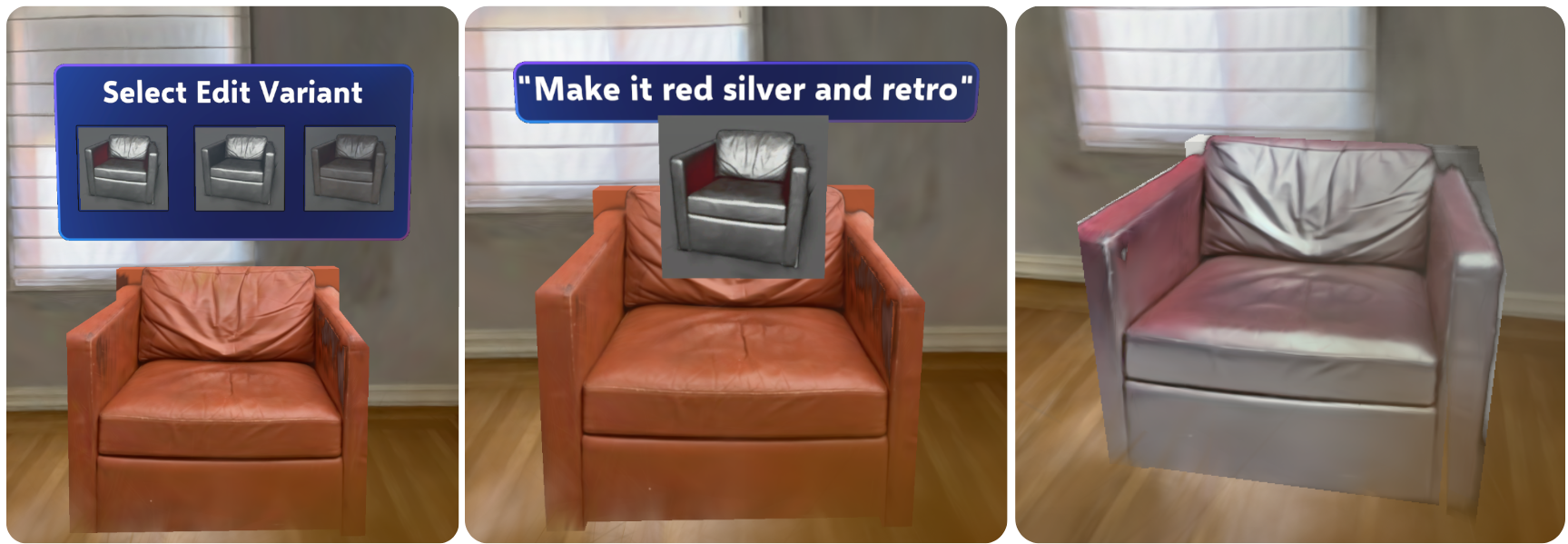}
\caption{\textbf{Radiance Field Object Editing with preview:} (Left) Edit variants are presented to a user. (Center) Displaying selected edit preview as a spatial annotation. (Right) Fully processed 3D edit replaces the original}
\label{fig:objEdit}
\Description{3D chair asset is placed in living room space, with three visual options for modified chair assets. Instructions hover above to “Select Edit Variant”. Upon selection, a single 2D sample modification image is displayed with the edit instruction prompt. Final image depicts a fully modified 3D asset, in style of instruction and 2D image selected.}
\end{figure}

\subsubsection{Edit radiance field objects via prompting}
Radiance field objects can be given stylistic or basic structural edits by pointing at an object and speaking an instruction, e.g. ``Make this chair chrome and futuristic.'' See \autoref{fig:objEdit}. A render of the object is given to the Instruct-Pix2Pix module, which applies the instruction to show as a 2D preview of the edit. We chose to use Instruct-Pix2Pix to preview this edit since it is a 2D equivalent of the 3D edit modules we use. Users can select from three edit variants, which will be applied for the final 3D object edit. Users can re-prompt edit instructions to quickly iterate and preview before running a time consuming full 3D edit. Edits take approximately 10 seconds to generate previews. Objects can be duplicated, re-edited, or deleted.


\begin{figure}[h]
\centering
\includegraphics[width=\linewidth]{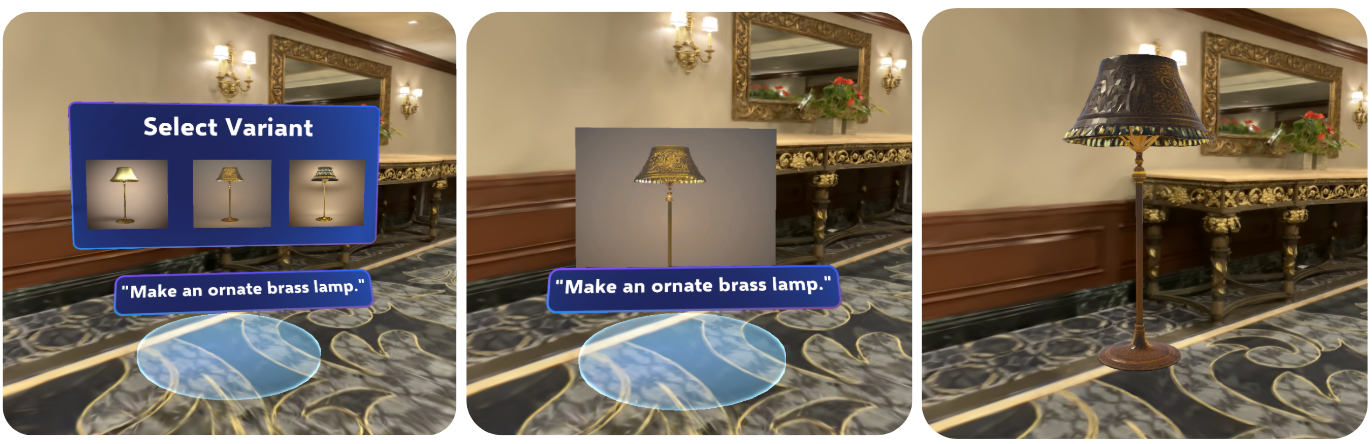}
\caption{\textbf{Object Generation via Prompting:} (Left) Object generation variations from speech input. (Center) Displaying selected generation preview as a spatial annotation. (Right) Fully processed 3D generation in the scene.}
\label{fig:objPromptGen}
\Description{Floating selection platform is placed on the floor in a hotel hallway scene. User prompts to “Make an ornate brass lamp”, and is presented with three 2D image options with instructions to “Select Variant”. User selects a single image option, and thereafter a 3D asset is generated and placed in scene matching the image variant selection.}
\end{figure}

\subsubsection{Generate objects via prompting}
Users can generate objects by pointing at the ground and speaking the prompt of the object they want to create (\autoref{fig:objPromptGen}). This sends an API call to the 3D generative module that includes Shap-E~\cite{jun2023shapE}, which generates a low fidelity mesh and render, and the render is stylized using depth conditioned ControlNet~\cite{zhang2023adding} with the original spoken prompt added with tags for photorealistic outputs. Optionally, the object generation and image stylization module can be themed to the scene through in-painting and masking methods. The user can select from three stylized 2D image variants of the object which are generated via ControlNet running with different random seeds, allowing greater variation in texture and style and slight changes in overall shape. Generations take approximately 15 seconds to generate previews. During an offline process, the full fidelity 3D objects are generated, exported as textured meshes, and placed in the scene.

\begin{figure}[h]
\centering
\includegraphics[width=\linewidth]{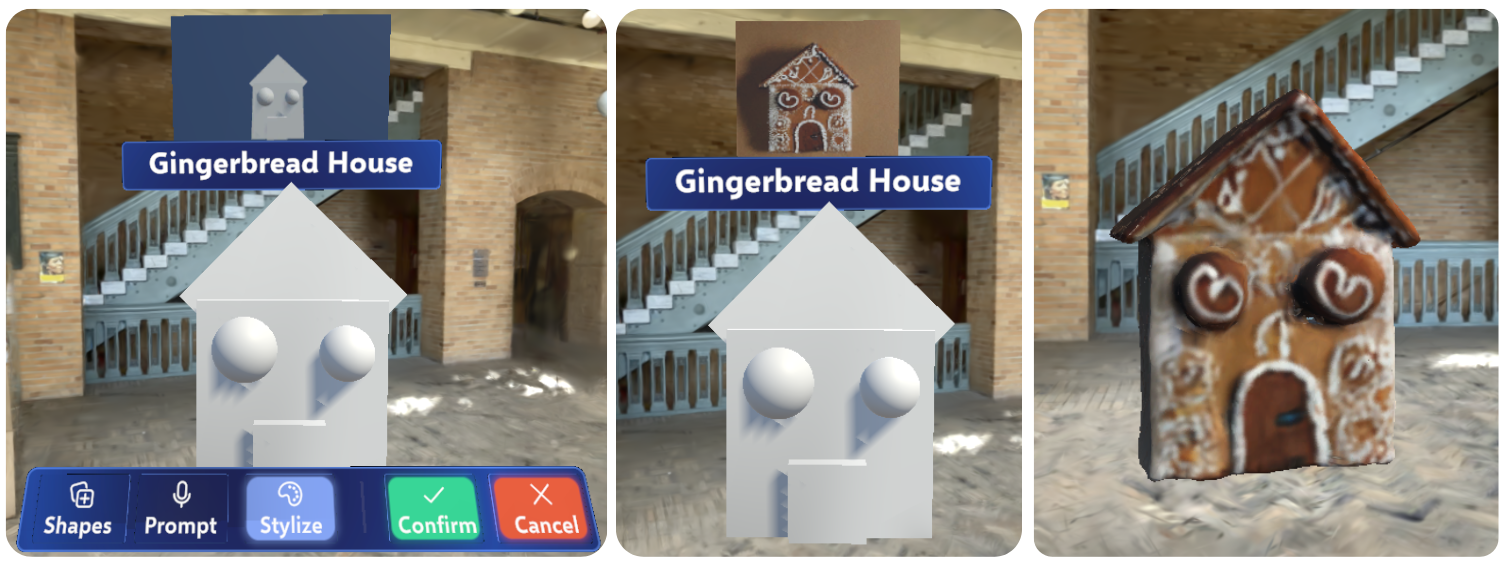}
\caption{\textbf{Object Generation via Sculpting:} (Left) Sculpting toolkit to create primitive shape arrangement (Center) Displaying stylized sculpted object preview as a spatial annotation. (Right) Fully processed 3D generation in the scene.}
\label{fig:objSculptGen}
\Description{An untextured 3D asset composed of basic shapes is depicted on the floor in a large indoor 3D space, resembling gingerbread house style. Horizontal menu bar appears with options “Shapes, Prompt, Stylize, Confirm, Cancel”. User prompts “Gingerbread House” and 2D image preview of colored and stylized gingerbread house is displayed above the untextured 3D asset. Finally, the 3D asset adopts a fully textured gingerbread house in the scene.}
\end{figure}

\subsubsection{Sculpt then stylize objects}
Alternatively, users can generate new objects by creating an arrangement of basic 3D primitives (i.e., spheres, cubes, and cylinders) (\autoref{fig:objSculptGen}). A limited set of tools are provided to position, rotate, and uniformly scale the shapes to "sculpt" a coarse low fidelity 3D shape. To make a higher fidelity textured model, the tool then takes a snapshot of this arrangement, and stylizes the render with depth conditioned ControlNet conditioned with a user-given prompt of the intended object. The stylization with ControlNet (or a similar depth conditioned re-stylization image model) essentially takes in an input image and text prompt, and re-stylizes the image by changing its texture improving its fidelity, even slightly refining its shape without changing too much of its general structure.
Once the user confirms the stylized and sculpted 2D generation, the full fidelity 3D object is generated offline with an image-to-3D generative model, exported as a textured mesh, and placed in the scene. We decided to provide this method of object generation to give users finer control over the generation compared with a text prompt and another method to translate low fidelity inputs to high fidelity outputs in 3D (translating existing recent 2D sketch to image systems to 3D editors) which is a new interaction possible with generative AI models, not found in traditional 3D tools like Unity or Blender. This implementation limits users to create shape arrangements but extensions could include more sculpting tools or free form 3D drawing.

\subsection{Radiance Field Objects}
Users can add radiance field objects or environments into the scene from a library of pre-captured objects. In our implementation we use gaussian splatting objects we captured ourselves which are trained and processed in Nerfstudio and Luma AI. Captures are pre-processed by segmenting individual objects and imported into Unity as ply files. To make them selectable and add collision physics in Unity, we have a mesh inside the radiance field object that enables interactivity and anchors the Gaussian Splatting object to the mesh.

\subsection{Proxy representations: Labels and Previews}
Proxy representations are intended to help users see the impact of their editing operations in real time. There are two types of proxy representations: labels and image previews. \autoref{fig:objEdit} (center) and \autoref{fig:objPromptGen} (center) show the labels and image previews. The labels show the prompts users have spoken aloud as commands to the generative AI modules (e.g., ``make the sofa blue''). The image previews show 2D versions of the anticipated generation. These image previews are generated using Instruct-Pix2Pix which is the underlying 2D image editing system used for the 3D radiance field editing system, Instruct-NeRF2NeRF.

Both the labels and image previews are associated with radiance field objects in the scene. This is done through a spatial annotation framework we developed. The framework logs each object's positions, object type, generative AI prompt, and image preview to a JSON file used for 3D generation and replacement, which we discuss next.

Good proxies should be fast to generate and accurate in previewing the final object. Dreamcrafter uses 2D image proxies because existing 3D object editing and generation pipelines use 2D images under the hood. For example, Instruct-GS2GS uses Instruct-Pix2Pix to first generate a 2D image from a natural language prompt and then transform the 3D scene into its edited version guided by the 2D images. By accessing the generated Instruct-Pix2Pix 2D image as the proxy in seconds, Dreamcrafter is able to show a preview quickly and, critically, by design, ensure that the 2D image is an accurate proxy of the 3D object. 

In other words, Dreamcrafter’s approach to generating proxy representations contributes a generalizable template for leveraging intermediate representations of high-latency 3D operations as proxies.

\begin{figure}[h]
\centering
 \resizebox{1.0\linewidth}{!}{%
      \includegraphics{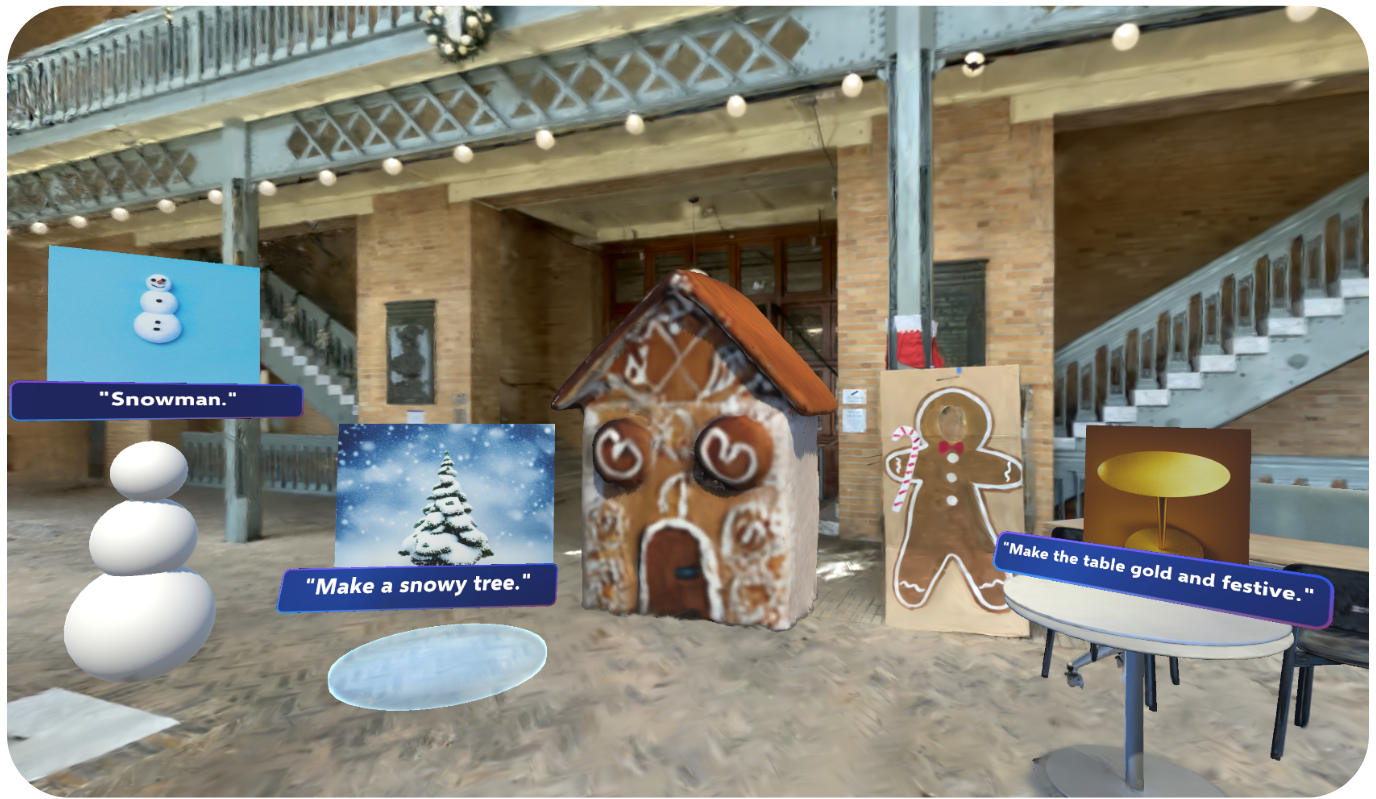}
}
\caption{\textbf{Spatial annotations:} tags and proxy representations are placed over the radiance field objects and generated objects with given instructions and preview generations.}
\label{fig:spatialAnnotations}
\Description{The scene depicts a floor of a large indoor space. There are various 3D assets rendered around the environment, including an untextured snowman, a generative platform represented by a disc, a fully textured gingerbread house, and a circular table with chairs behind. The snowman is accompanied by a spatial edit instruction “Snowman” and 2D render of a snowman. The platform has a spatial tag “Make a snowy tree” along with a 2D render image. The gingerbread house is fully rendered and stylized. The table is textured, and has a spatial edit to “make it gold and festive” accompanied by a 2D image depicting the new style to implement.}
\end{figure}

\subsection{Modular System Design using Generative AI modules}
Dreamcrafter's Unity client offers a modular interface to multiple plug-and-play modules for real-time interactive and offline processing tasks. The system is designed to easily update to newer iterations of these generative AI models, which are commonly developed due to rapid interest in this field.
\subsubsection{Online Processing Modules}
A set of generative modules are used to create rapid previews visible in the VR scene. Radiance field object editing tasks use Instruct-Pix2Pix, an intermediate model used for the full 3D edit which runs in 15 seconds. The object generation via prompting instruction uses a text to 3D module, Shap-E, to generate a low fidelity mesh and NeRF render. This render is then stylized with ControlNet conditioned on edges and the same text prompt to create three 2D preview variant generations. We use Shap-E since it creates a render of a single object and an object centric generation than a regular text to image model, and provides a close approximation of using a more detailed text to 3D system. Object generation via sculpting displays a 2D preview generated via ControlNet conditioned on depth and a snapshot of the arranged 3D primitives. The sculpted arrangement acts as a 3D proxy representation. 

\subsubsection{Offline Processing}
Using the JSON log output from the spatial annotation system, Dreamcrafter makes instruction and tool specific API calls for each generative AI module. A Python broker server receives a server message from the Unity project and forwards instruction parameters (e.g., instruction type, text prompt, image input) to the specified module. \autoref{fig:teaser} shows an overview of the system architecture. 
Object generation uses a 3D generative module Shap-E, and a 2D image stylization module ControlNet and Stable Diffusion. The full object 3D generations use 2D-image to 3D-model models such as LGM~\cite{tang2024lgm}, GRM~\cite{xu2024grm} or any text-to-3D based system. The final 3D object edits are done using Instruct-GS2GS \cite{igs2gs} for Gaussian Splatting objects. 
The modules are exchangeable and can be implemented to use updated AI models.
After the edited objects are added to the scene, users can repeat the process and edit the scene again, creating an iterative design process. 

\subsection{Scene Outputs}
\subsubsection{3D Scenes}
After offline processing, fully edited scenes can be viewed as a 3D Unity scene composed of radiance field objects and meshes. Optionally, training images can be captured of the scene to create a radiance field of the entire scene.

\subsubsection{Magic Camera}
\begin{figure*}[h]
\centering
 \resizebox{1.0\linewidth}{!}{%
      \includegraphics{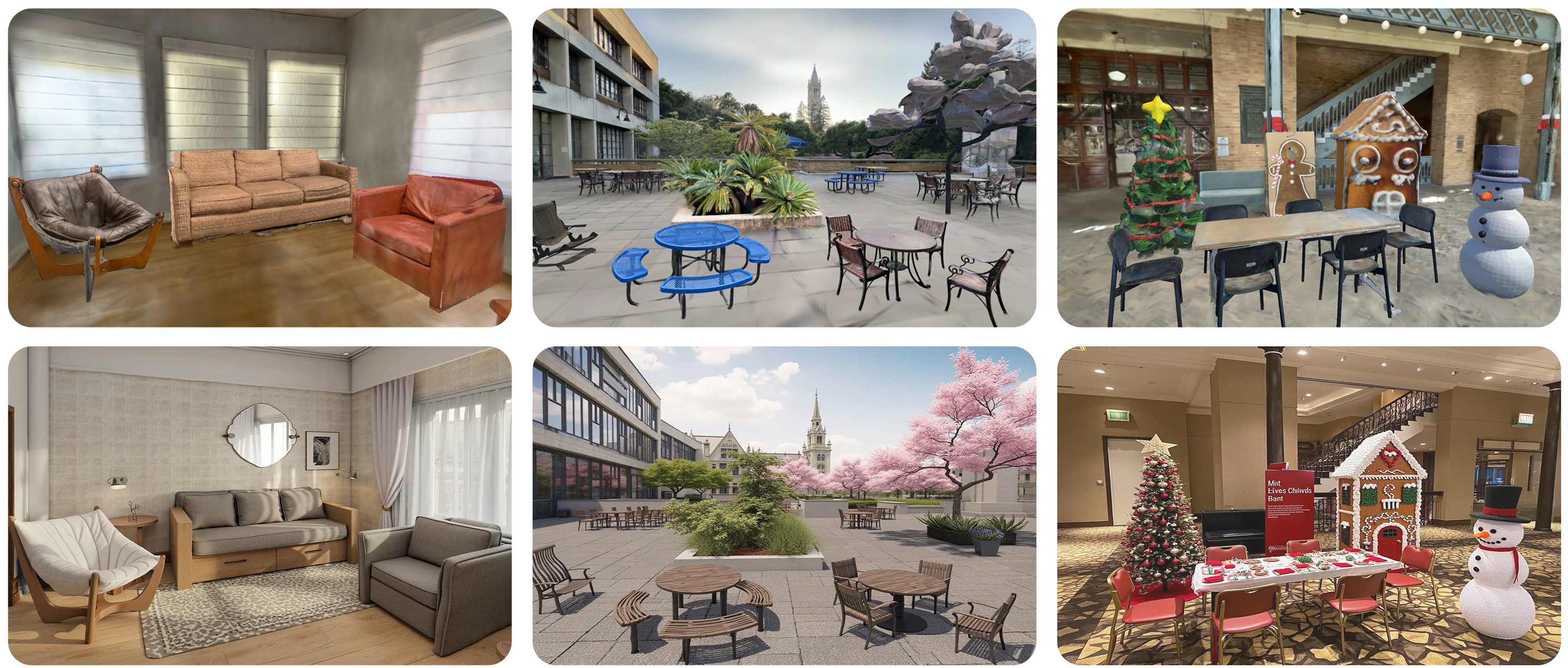}
}
\caption{\textbf{Magic Camera Outputs:} (Top row) Scene inputs from the virtual Unity camera (Bottom row) Stylized image outputs given the prompts: (Left) "realistic apartment living room", (Center) "realistic outdoor patio in university with cherry blossom tree, chairs, benches, plants", (Right) "a holiday party in a hotel atrium, snowman, Christmas, gingerbread house". The image stylization model generates stylistically consistent and coherent photorealistic image outputs from the input scenes of composed radiance fields and 3D models.}
\label{fig:magicCamFig}
\Description{Figure 9: Three sets of inputs and outputs for the Magic Camera image re-stylizing system. The top row shows the scene inputs: (Left) A living room scene with a sofa and chairs placed in a 3D scene, (Center) An outdoor deck next to a building with benches, chairs, plants, and a 3D model of a cherry blossom tree and chair. (Right) A holiday party scene composed of tables, chairs, a 3D models of a Christmas tree, a snowman, and a gingerbread house in a building atrium. The bottom row shows the coherent photorealistic outputs: (Left) A realistic living room scene with correct lighting and shadows. (Center) An outdoor deck next to a building with stylistically consistent benches and chairs with plants and a realistic tree. (Right) A stylized version of the same holiday party scene in a photorealistic style with realistic lighting and shading. }
\end{figure*}

Users can position a virtual camera, we call the Magic Camera, which stylizes a snapshot of a view of the scene given a prompt through an image re-stylizer (like ControlNet or FLUX.1 Depth \cite{flux}) via multi-modal input (image + text prompt). The resulting stylization gives a coherent and realistic composition of the scene based on the content and arrangement of objects, analogous to rendering a frame in a traditional 3D editor. The Magic Camera is implemented as a virtual camera in the Unity scene which the user can position and preview and enter text prompts via a floating panel. A user can select to capture a snapshot which sends a render of the Unity scene to the image re-stylizer ControlNet module and the output stylized render is shown on the panel (takes about 15 seconds). This stylizes all objects in the image snapshot and can add additional detail to lower fidelity generated objects in the render. \markupRed{We observed that the image stylization models were able to implicitly detect the kind of objects when stylizing based on their depth, and by mentioning the objects in the prompt, the model correctly stylized the objects. In Figure \ref{fig:magicCamFig}, just using the prompt "realistic living room" and inputting the image of the room, the model could detect the sofas and windows and accordingly stylize them correctly. Likewise, for the holiday party scene, by inputting the objects in the prompt (snowman, gingerbread house) the model correctly stylized the objects based on the depth map of the input image.} In its current implementation, the Magic Camera is limited to capturing only a single view and supports only a text prompt with an image as input into an image re-stylizer. However, it could be extended to support other multi-modal specialized prompts and models such as camera trajectory or an additional style reference such as for generative video models.

\begin{figure}[h]
\centering
 \resizebox{1.0\linewidth}{!}{%
      \includegraphics{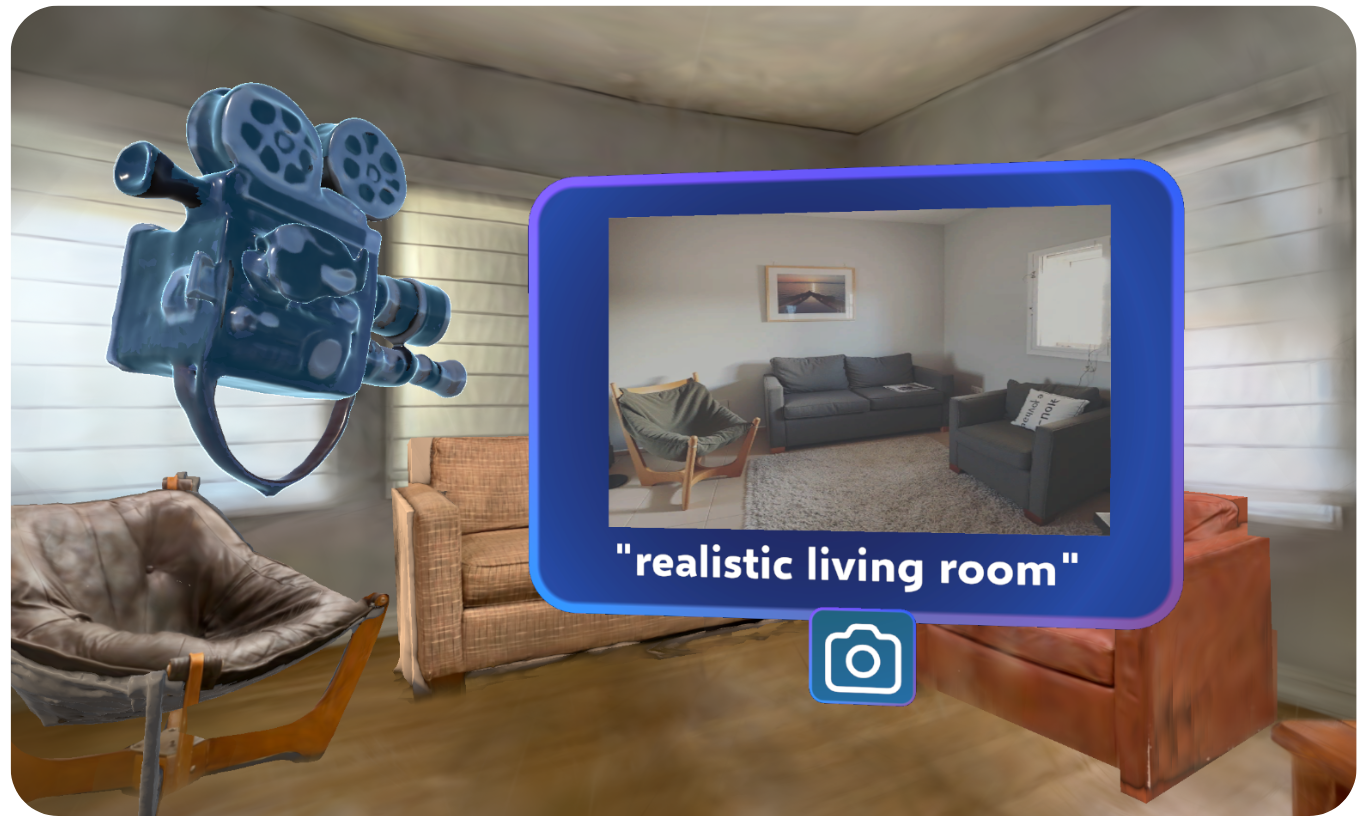}
}
\caption{\textbf{Magic Camera:} The virtual camera can be repositioned with a panel previewing the stylized output from the Magic Camera with the given text prompt.}
\label{fig:magicCamPanelFig}
\Description{A living room scene with a blue holographic camera and a panel with a preview of a stylized realistic render from the magic camera of its perspective with the text caption "realistic living room" on the panel.}
\end{figure}

This feature can be extended to act as a method of controllability in AI generated images or video by using This 2D image output as input to an image-to-video model. The magic camera output image could potentially be used as input into an image-to-3D scene system \cite{gao2024cat3d} which would generate a 3DGS scene, editable in Dreamcrafter. This could create an iterative design process where a user could create a general layout of the objects and positions in the scene, and can use the Magic Camera to stylize it, and then iteratively edit the 3D scene. The Magic Camera acts as a gateway to other generative modalities and outputs such as video or 3D scene generation and allows Dreamcrafter to act as a spatial interface for video generation.

\subsection{Additional implementation details}
Dreamcrafter is implemented in Unity using the Unity XR toolkit and MRTK plugins for VR support. Gaussian Splats were rendered using an open source Unity Gaussian Splatting viewer \cite{unitygs}, and example splats were trained using the Nerfstudio splatfacto model and Luma AI. The Unity app interfaces with the online generative modules using a sending calls from a C\# server to a python flask server which makes API calls to separate generative models with specified parameters using a modified Diffusers library ~\cite{von-platen-etal-2022-diffusers} for Shap-E and Instruct-Pix2Pix. The ControlNet module is run on the Stable Diffusion WebUI~\cite{sdwebui}. All generative models were run locally. We used Stable Diffusion v1-5 for proxy generations. The radiance field objects (gaussian splats) in the scene are photorealistic and edits to these are photorealistic. However, newly created objects using generative AI models are not always photorealistic due to limitations in current 3D generative models. In the future, as these libraries improve, the system could rely on newer libraries. This is possible because our system is modular.

\section{Evaluation}


Two research questions motivated the evaluation: 
\begin{enumerate}
    \item \textbf{RQ1 - Levels of control.} How do users want control over scene
    edits? Specifically, when do they choose to generate objects via
    prompting or sculpting? Why?
    \item \textbf{RQ2 - Proxy representations.}  What are users' reactions to the proxy representations? Are they sufficient for envisioning final scene edits?
\end{enumerate}

\subsection{Study Design and Procedure}
After participants gave informed consent, the researchers walked participants
through a tutorial introducing the interactions for editing and creating objects
in Dreamcrafter. The tutorial took approximately 30 minutes. Once participants practiced and expressed feeling comfortable performing the interactions, they were presented with the scenario of designing a 3D environment for a winter holiday party. They were asked to complete the following tasks: 

\begin{itemize}
    \item \textbf{Dining area for six.} Participants set up a dining area for six people. The 3D environment was already populated with a couple of tables and a chair that participants could duplicate or edit.
    \item \textbf{Photo area for party guests.} Participants decorated an open area for taking pictures. The task was to create a North Pole scene by considering snowmen, elves, or trees. 
    \item \textbf{Gingerbread house.} Participants created a gingerbread house with two windows and one door. 
    \item \textbf{Unstructured editing.} At the very end, participants were given five minutes for free-form editing where they could revisit any of the tasks above as they edited the scene to their liking.
\end{itemize}

We designed the tasks such that they required a range of editing and creating
operations, where different modalities would likely shine and showcase the flexibility of tools supported. The dining area task was the most scaffolded, with relevant objects
populating the scene already and a small object library of radiance field objects was given. We anticipated that this would encourage
participants to edit the existing radiance field objects or add relevant objects from the object library. The photo area
was more open-ended with opportunities to place objects and generate new
ones via prompting or sculpting in an open area of the room. We use this task to examine when users decide to prompt or sculpt and the benefits of the proxy representations for scene composition. The gingerbread house was the most specific
task, likely requiring a significant amount of control. For all the tasks,
participants were encouraged to use any interaction as they saw fit. 

Upon completing the tasks, participants completed an exit survey and interview. 
In total, the study lasted approximately 90 minutes. The participants used Dreamcrafter in a Meta Quest 3 with PC link.

\subsection{Participants} Participants were recruited via word-of-mouth through school tech-related Slack channels (from VR courses and clubs),
newsletters, and mailing lists. Participants self-reported having relatively little experience in VR (median=2/5). Four of the seven participants had prior experience with 3D tools (Unity or Blender), and two participants had prior experience with creative generative AI tools. Participants were
compensated \$35 for their time. 

\subsection{Measures and Analysis}
For each task, we recorded and analyzed videos for how participants manipulated objects (i.e., editing vs. creating; prompting vs. sculpting) and why. We also thematically analyzed their open-ended survey questions and interview responses.

\subsection{Results}
Overall, participants reported that Dreamcrafter helped them edit the scene as they wished [P2, P4, P6, P7]. P5 expressed how the scene they created using Dreamcrafter was ``not what [they] thought but more interesting.'' due to the sometimes unexpected results from the generative models.

\subsubsection{RQ1: Levels of control}
Overall, participants rated their success in achieving their desired edits highly (Dining area: median=5/7, Picture area: median=5/7, Gingerbread house: median=4/7). For all tasks, participants more frequently generated objects using prompting instead of sculpting. Four out of seven participants used a mixture of prompting and sculpting across the study tasks (\autoref{tab:evalCounts}). Three even used both prompting and sculpting within the same task. For example, P1 created most of the gingerbread house via sculpting but then wanted to augment it with prompt-generated windows. 

\begin{table}[t]
\caption{\textbf{Evaluation: Different levels of control used.}}
\label{tab:evalCounts}
\centering
\small
\begin{tabular}{l|lll}
\multicolumn{4}{p{245pt}}{The number of objects created using each approach are in parentheses. 
    Participants used a combination of editing existing objects, creating objects via prompting, and creating objects through sculpting throughout the tasks. Four out of seven participants  used a combination of prompting and sculpting throughout the study, including sometimes for the same task. While the majority of participants created the majority of objects via prompting alone, participants reported gravitating towards sculpting to control generation.
}\\
\hline
\colH{ID} & \colH{Dining area}          & \colH{Photo area}           & \colH{Gingerbread house} \\
   \hline
P1 & Edit (2) & Prompt (1) & Prompt (1), Sculpt (1) \\
P2 & Prompt (2) & Prompt (3) & Prompt (1) \\
P3 & Edit (2) & Prompt (3) & Prompt (3) \\
P4 & Edit (1) & Prompt (3), Sculpt (1) & Sculpt (1) \\
P5 & Prompt (4) & Prompt (6) & Prompt (1), Sculpt (6) \\
P6 & Edit (2) & Prompt (3) & Sculpt (1) \\
P7 & Edit (2), Prompt (1) & Prompt (4) & Prompt (1) \\
\end{tabular}
\Description{Counts of the levels of control (i.e., Editing, Creating via prompting, Creating via sculpting) participants used for each task in the evaluation.}
\end{table}

When asked why they chose to create objects via prompting, participants explained that prompting was easier to use [P2, P3, P4, P5, P7]. Prompting helped them ``save time'' [P1], required less active user involvement [P2], and resulted in ``more polished'' results [P3]. P4, explained, \shortquote{The prompting tool did make it extremely easy to take what I am thinking and make a relatively accurate depiction.} 

Participants had mixed opinions on how well prompting served their goals when they had specific details in mind. P1 and P6 explained that they preferred prompting over sculpting
depending on \shortquote{typically how complicated I expected the object to be} [P6]. At the same time, P4 reported \shortquote{[the generated 2D proxy representation] sometimes fell short in some minor details of what was described in the prompt.} 

In contrast to prompting, participants reported feeling they had more control when sculpting then stylizing objects [P1, P4, P5]. P4 explained, \shortquote{if I had an idea in my head that I know how I wanted it to look like...it kind of had a little more restriction what the AI used to create versus the prompting}. When asked when they chose to sculpt, P1 and P5 explained that they preferred sculpting large-scale objects, such as the gingerbread house. At the same time, most participants, including P7 who did not use sculpting, wanted to have access to more shapes [P4, P5, P6, P7] and finer grained object manipulation [P2, P4, P6, P7], suggesting that sculpting may ultimately be more desirable than we saw in our study. 
 
\subsubsection{RQ2: Proxy representations}
Six out of seven participants primarily relied on the image previews to get a sense of the scene's overall composition [P1, P2, P3, P4, P6, P7]. For example, P1 described how the previews were ``helpful to put stuff around and see how it works for each other.'' Similarly, P3 remarked how each preview ``helps for arrangement in the space.'' 

Participants also reported that the image previews helped them visualize individual objects [P4, P5, P6]. For instance, P6 said ``It was easy to create an object that was somewhat close to what I was envisioning based on the preview it generated.''. Participants would re-prompt once or multiple times if they weren't satisfied with their initial generation for editing or object creation and would make their prompts more refined on their intent.

Despite reporting that the previews were helpful for scene composition and object styling, when asked how sure they were about how the final scene would look, P1, P2, and P6 reported feeling unsure, rating their certainty at a 1 or 2 on the five-point scale. The median score across all participants was a 3 out of 5. P5, who found previews helpful for envisioning individual objects but not the entire scene, pointed out a key limitation of the previews was that size information was lost: ``Some preview of the size an object would take would be useful for just the prompting / not sculpting part.'' Therefore, proxy representations, while helpful for drafting scenes and objects, are incomplete for fine-grained scene layout and detailed editing. 

Participants first envisioned and then described to the researchers a scene based on the task instruction, then generated objects, and finally positioned them. Some participants had a particular style in mind (P1, P5) and tended to generate/edit objects to achieve this style. Four participants (P1, P4, P5, P6) chose to use sculpting for the gingerbread house construction task to control generated details (e.g., placing two spherical windows above a rectangular door). During the dining scene, most participants opted to use the existing radiance field objects for tables and chairs. Four participants further stylized the existing objects to be consistent with their desired theme (e.g., Game of Thrones-esque).
\subsubsection{System Limitations and Strengths}
A primary limitation was the scene's physics. For six of the seven participants, rotating and arranging objects in the scene were difficult [P2, P3, P4, P5, P6, P7]. For example, when editing the dining area, P2 expressed \shortquote{When chairs would fall over, it was very hard to put them back up. Also, if I wanted to rotate or move the chairs they would tend to change size, so by the end most of the chairs were all different sizes.}

A noticeable limitation during the tasks was that the sculpting tool was sometimes difficult to use effectively. It took some time for the users to create the desired arrangement of shapes and users wanted additional familiar functionality present in most other systems (duplication, grouping, deletion). This difficulty may have influenced their experience and affected the accuracy of the comparison with prompting, which was much easier to use. 

Another important limitation was inaccurate speech recognition, which became a major burden for users relying on prompting [P1, P2, P3]. Despite this, most participants relied on prompting for setting up the picture area and gingerbread house, so we would expect that improved speech recognition would lead to more reliance on prompting. Related, because the system had a five second speech detection window, P5 expressed wanting the system to give them more time to express all the details they had in mind. In addition, the text-to-image models sometimes provided unexpected or low quality generations which required users to re-prompt the system multiple times in line with limitations of current image generation models, which some participants were aware of. 

Other technical challenges that participants reported were feedback time while waiting for Stable Diffusion results [P1, P5], awkward VR controller mappings [P6, P7], discomfort in VR [P2, P6].

Users could re-edit radiance field objects to have them match their intended style, and typically iterated a few times to refine their prompt. The three generated previews helped users choose their selection.
Participants expressed their reaction to the Magic Camera as "cool" [P1, P6, P7] and "useful" [P3], and some found the generations interesting [P5, P6] and "more thematic" [P5]. However, some expressed that  the generations were sometimes  "confusing" [P3] or not what they "envisioned" [P5] and that it "does not always align with what is presented" [P6].

Despite challenges with object manipulation and speech recognition, all participants expressed wanting to use Dreamcrafter in the future for a myriad of reasons: interior design [P1, P3, P6, P7], ``my creative side'' [P1],  CAD in engineering [P4], and video game design [P5]. P2 preferred to use a non-VR version. For P5, P6, and P7, generating objects via prompting was the best part of the system. This suggests that even with user experience issues, providing multiple forms of user control, proxy representations, and access to generative AI modules were desirable for diverse spatial computing applications and users. 

\subsection{Revisions to System}
\begin{figure}[h]
\centering
 \resizebox{0.9\linewidth}{!}{%
      \includegraphics{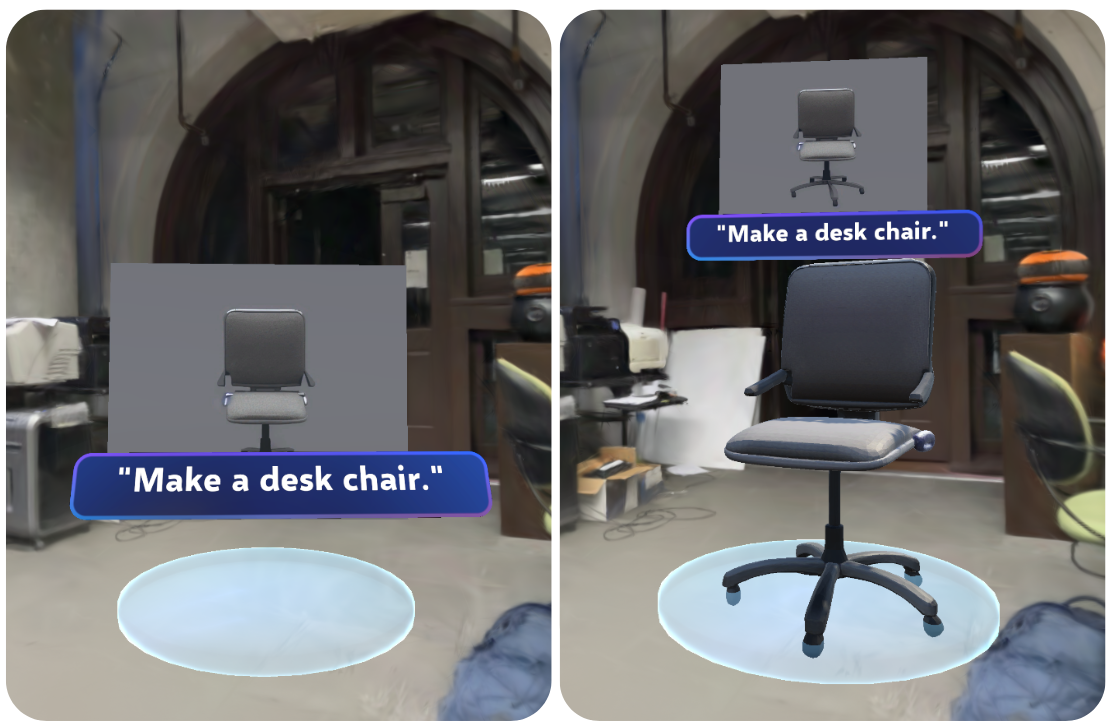}
}
\caption{\textbf{3D and 2D Proxy Representations:} (Left) New 3D proxy showing a low fidelity mesh preview (Right) Original 2D proxy representation with image preview}
\label{fig:figproxy}
\Description{(Left) Shows a 2D image plane of a stylized table placed in the scene. (Right) Shows a desk chair 3D model with a label “Make a desk chair” and a 2D image of a stylized chair over it.}
\end{figure}

Based on our preliminary user study, we updated the system to address user concerns and improve existing features. Based on feedback regarding the 2D proxies (specifically from P5's comments on scene composition), we implemented 3D proxy representations for object generation. This method imports the intermediate low-fidelity mesh which can then be placed and scaled in the scene and give the users a better sense of the object placement, as well as work better with the Magic Camera by providing a reference for an object. We show a comparison between the original 2D and new 3D proxy in \autoref{fig:figproxy}.

To prevent the need for users to over-explain a prompt to get a detailed stylistic generation as we observed in our study, we also experimented with the concept of generating more detailed prompts with additional scene specific context by appending specific keywords and prompts to the object generation prompt to make the generation stylistically consistent with the scene objects. We use GPT-4o with vision and prompt \shortquote{Act as an AI world building assistant and given this is the view from my vr headset, I want to use speech to generate a new object in this space and given the prompt, a ML model creates a 3D object out of it. Given a prompt and image, I want you to make sure the object appears stylistically similar to the scene shown in the image and other objects by adding additional keywords to the prompt to describe the color, material, and other structural details. I will tell you a prompt and give an image and you will give the slightly longer version of the prompt with the detail to make it stylistically consistent.} Given an images of the scene and a very short prompt of the object, it adds descriptions of materials and colors so the generated object matches with other objects in the scene. We experimented with this pipeline independently, but have not integrated it in the full system yet.

\section{Discussion}
We investigate how to incorporate the benefits of real-time, immersive editing and the advantages of high-level scene editing using generative AI. We develop and evaluate the Dreamcrafter system, which provides a modular architecture for generative AI algorithms, offers different levels of interactive control, and leverages proxy representations to show previews of high-latency edits to radiance field objects. 

Through a preliminary first-use study, we find that users, including those without VR or scene editing experience, find the direct manipulation (sculpting) and natural-language based (prompting) interactions useful for editing and creating objects. Most use a mixture of both interactions. Sculpting objects and then stylizing them with generative AI helps participants feel they have more control over the generation process. Yet, participants create more objects using only natural language prompts. This is not surprising given the relative speed with which generative AI models can create object proxies (previews) and being given a selection of variants which reduces the chance of selecting unfavorable generations. Interestingly, despite the control direct manipulation affords them, participants preferred generative AI-based object creation over sculpting when they had very specific details for what they wanted objects to look like. These findings suggest that sculpting may be useful for giving the general shape of an object while prompting is useful for its specifics. Both sculpting and prompting appear to serve different purposes in users' design processes, so supporting both forms of control is necessary for scene editing tools to support a diversity of creative paths and styles~\cite{resnick2005design}. Using generative AI in these workflows helps automate intensive scene creation tasks such as 3D modeling and texturing, but at the cost of finer grain control and unpredictable quality of generations. Such systems like Dreamcrafter can limit functionality and workflows to the system or model capabilities which could influence the user's artistic expression.
As observed in the study, participants appreciated the generations and usually found them interesting but left it to chance if the generations were of high quality or of original intent. We noticed that some preferred to see what the system created and used it as inspiration for either re-prompting or creating new objects while others wanted fine-grained control which minimized the chance of unintentional generations. Supporting both creative processes is important for lowering the barriers to 3D scene creation in the future.



Furthermore, participants found Dreamcrafter's 2D proxy representations of high-latency 3D object editing and creating operations useful for editing 3D scenes. This suggests the importance of realtime feedback for spatial computing tasks. This also suggests that leveraging 2D generation for 3D scenes may be a promising path forward for providing realtime feedback. We find our study opens future directions for research in further evaluating using immersive editing, adding additional multi-modal interactions, and measuring cognitive load of participants. Additionally, providing both text and image proxy representations may be especially important for future semantic, generative AI-based scene editing systems. We believe Dreamcrafter's proxy representations, though designed for image and 3D systems, can be generalized for other visual outputs from generative models such as video. We hope that using real-time proxies could be a contribution to graphics research in enabling real-time interactions for longer processes.

Overall, in Dreamcrafter, we explore not only the feasibility but also the benefits of providing both rapid direct manipulation and high-level instruction-based editing support in 3D scene editing. Through varying levels of control and proxy representations, Dreamcrafter is a step towards continuing to lower the barriers to 3D scene editing, especially for emerging graphical representations such as NeRFs and Gaussian Splats.

\paragraph{Applicability to future user interfaces for generative models}
We believe that Dreamcrafter could also act as a world creation or staging tool for other generative AI design systems for 2D or video output, we call spatial prompting. With the inception of recent advances in generative models, we believe that there is a desire for more visual interfaces to image/video generative models in consumer applications. A system we explored during the project's development was using the Magic Camera to pre-visualize stylized scenes through ControlNet and Stable Diffusion based on the construction of a scene of only primitive objects, created and arranged within the VR interface. Even with minimal object detail (e.g., cubes as a couch), the system produced highly stylized, recognizable scenes and objects based on a single global scene prompt. Future improvements could involve tagging objects for individual stylization and converting 2D renders into 3D scenes.
Dreamcrafter could serve as an early exploration into spatial prompting systems that offer more control for 2D/3D/video scene generation systems beyond limited text prompting interfaces which are currently in SOTA consumer applications. Scenes and objects could be designed at a higher abstraction level through primitive objects. These lower fidelity representations are much easier to design and iterate, and can offer a variety of different higher fidelity generations from the given arrangement of primitives using methods from stable diffusion and ControlNet generalized to 3D objects and scenes. These lower fidelity proxy representations, optionally paired with semantic information like text prompts, could help add controllability in 3D scene generations. Arrangements of proxy representations could also be sourced from other mediums such as images or videos of arrangements of physical objects or gestures/motion from users, potentially using an LLM to interpret vague instructions. In our implementation of Dreamcrafter, we primarily use gaussian splats as the radiance field 3D representation, but we anticipate that our interactions and system are generalizable to future implicit 3D representations. In the case of virtual production and pre-visualization, methods discussed above could be used to create a system that enables users to create low fidelity approximations of scenes, movement of objects, and camera movement as input modalities to generate a stylized high fidelity output from a video diffusion model. As described in Sora’s technical report \cite{videoworldsimulators2024}, video diffusion models may have the potential to generate large scale 3D scenes and virtual worlds. These could be also edited through methods in Dreamcrafter discussed above or used to complete or extend 3D scenes. These methods could leverage all capabilities of the editing and generation systems presented in Dreamcrafter for world building systems. 
We envision Dreamcrafter to be a system for editing and assisting in the creation of realistic environments and worlds for future 3D representations. In order to approach this goal with today's systems and technology we create an immersive gaussian splatting editor with multimodal generative models. Given that multiple startups like WorldLabs, Luma AI, Open AI, Runway ML are exploring world building systems in the form of video and generative 3D systems, we hope that our work can inform the development of these future systems and graphics/computer vision research for more broadly world generation systems.




\section{Limitations and Future Work}
There are a few limitations to this work that offer opportunities for future work.

\paragraph{Global scene editing.}
Dreamcrafter supports editing and creating individual radiance field objects within an environment. However, users may want to edit aspects of the underlying environment as they design their scenes. One way we have begun to explore this possibility is through developing functionality that allows users to take a snapshot of an environment from a fixed perspective and then stylize that snapshot, in a manner similar to how sculpted objects are stylized in Dreamcrafter currently. The resulting generation suggests a possible way to stylize the scene and all objects contained within it together. Ideally, users should be able to define the perspectives they take snapshots from and how they stylize the scene, perhaps even controlling which objects receive the global style treatment. 

\paragraph{Additional ways to control generation.} A key focus of future work should be the development of more intuitive ways to generate radiance field objects. For instance, rather than rely solely on voice commands, what if users could use Dreamcrafter with text or 2D/3D sketches/images as reference input or with multimodal input like gestures, which then get translated into or serve as generative AI prompts?
Incorporating voice commands for positioning like ``place the table next to the blue chair'' would make the system more user-friendly without having to manually place objects.

We anticipate that Dreamcrafter's modular design will help explore new interaction techniques. Dreamcrafter has separate modules for object generation, for using AI to create new objects, and spatial annotation, for placing objects in the scene. By separating these concerns, Dreamcrafter has the potential to evolve with not only new AI technologies but also new 3D representations (i.e., whatever may replace radiance fields for photorealistic rendering in the future). 
\paragraph{Even more rapid proxies.} 
While Dreamcrafter currently supports speech-to-text prompt labels and image previews, what might alternative proxies or intermediate proxies between 2D and 3D objects look like? For example, would users find 3D wireframe outlines just as useful as the 2D image previews? Furthermore, if users could stylize entire scenes, what would the appropriate proxy for the entire scene be? A sketch of the new alongside the old? Over the course of the project's development, recent image-to-3D models released which can run within seconds instead of minutes, and it could be possible to explore new proxy representations that take advantage of the faster inference while keeping the same foundation of the overall system.
\paragraph{Automatic Segmentation.} Dreamcrafter currently takes in input of full 3DGS and objects, however it currently is unable to edit objects that are fixed in the scene. To enable editing and placement of objects baked in existing scenes, having automatic semantic segmentation could be used to streamline the editing workflow, making it more efficient for users, without requiring manual segmentation. Users could be able to select regions of objects via segmentation or volumes in space to perform edits.

We believe that these avenues of future work can apply to future 3D editors and generative world building systems.



\section{Conclusion}

The idea behind Dreamcrafter is to use direct manipulation for spatial positioning and layout; and leverage generative AI for editing style and appearance of photorealistic objects. Because generative AI edits are unlikely to run in real-time, Dreamcrafter introduces rapid proxy representations, e.g. using a 2D diffusion model to create a stand-in image for a longer-running 3D generative task.  Dreamcrafter enables both 2D (image) and 3D output. In a first-use study, participants report feeling more in control of AI generation when they first sculpt objects before stylizing them with generative AI. Participants also report finding proxy representations useful for scene editing. We discuss how Dreamcrafter could help advise future work in world building systems.

\bibliographystyle{ACM-Reference-Format}
\bibliography{sample-base}

\end{document}